\documentclass[english,aps,pra,twocolumn,superscriptaddress]{revtex4-1}
\usepackage[T1]{fontenc}
\usepackage{graphicx}
\usepackage{babel}
\begin{document}

\title{Certification of high-dimensional entanglement and Einstein-Podolsky-Rosen
steering with cold atomic quantum memory}

\author{Micha\l{} D\k{a}browski }

\affiliation{Institute of Experimental Physics, Faculty of Physics, University
of Warsaw, Pasteura 5, 02-093 Warsaw, Poland}
\affiliation{Centre for Quantum Optical Technologies, Centre of New Technologies, University of Warsaw,  Banacha 2c, 02-097 Warsaw, Poland}

\author{Matuesz Mazelanik}
\email{mateusz.mazelanik@fuw.edu.pl}

\selectlanguage{english}%

\affiliation{Institute of Experimental Physics, Faculty of Physics, University
of Warsaw, Pasteura 5, 02-093 Warsaw, Poland}
\affiliation{Centre for Quantum Optical Technologies, Centre of New Technologies, University of Warsaw,  Banacha 2c, 02-097 Warsaw, Poland}

\author{Micha\l{} Parniak}

\affiliation{Institute of Experimental Physics, Faculty of Physics, University
of Warsaw, Pasteura 5, 02-093 Warsaw, Poland}
\affiliation{Centre for Quantum Optical Technologies, Centre of New Technologies, University of Warsaw,  Banacha 2c, 02-097 Warsaw, Poland}

\author{Adam Leszczy\'{n}ski}

\affiliation{Institute of Experimental Physics, Faculty of Physics, University
of Warsaw, Pasteura 5, 02-093 Warsaw, Poland}
\affiliation{Centre for Quantum Optical Technologies, Centre of New Technologies, University of Warsaw,  Banacha 2c, 02-097 Warsaw, Poland}

\author{Micha\l{} Lipka}

\affiliation{Institute of Experimental Physics, Faculty of Physics, University
of Warsaw, Pasteura 5, 02-093 Warsaw, Poland}
\affiliation{Centre for Quantum Optical Technologies, Centre of New Technologies, University of Warsaw,  Banacha 2c, 02-097 Warsaw, Poland}

\author{Wojciech Wasilewski}

\affiliation{Centre for Quantum Optical Technologies, Centre of New Technologies, University of Warsaw,  Banacha 2c, 02-097 Warsaw, Poland}
\begin{abstract}
The advent of complex, structured and high-dimensional entangled statesbring both new possibilities for experimental and theoretical scenarios as well as new challenges for generation and characterization of such states. In particular, spatially-structured photonic states offer applications in quantum imaging, information processing, and quantum key distribution. Here we experimentally generate a spatially entangled high-dimensional state composed of at least 10 Schmidt modes in a quantum memory setup and perform characterization using the entropic EPR-steering inequality, yielding genuine violation of $1.06\pm0.15$ bits. The entanglement of formation of at least $0.70\pm0.15$ ebits for the measured noisy state is certified using the entropic witness method. We point out and solve the difficulties in estimating the entropy, achieving characterization of the high-dimensional entangled state with highly undersampled data. Finally, the practical supremacy of the entropic EPR-steering witness over a variance-based witness is demonstrated for a wide class of states typical in an experimental scenario, giving prospects for EPR-steering demonstrations and applications in noisy systems or with lossy quantum channels.
\end{abstract}
\maketitle
\section{Introduction}
Entanglement is not only a central concept of the quantum theory \citep{Vertesi2010},
but also an essential resource in a variety of quantum-enhanced practical
protocols, such as secure key distribution \citep{Gisin2002}. Of particular
interest is the broad concept of high-dimensional (HD) entangled states,
for which the dimension of Hilbert space required to describe the
observed correlations reaches thousands \citep{Krenn29042014}. These
states offer an unprecedented informational capacity \citep{Dixon2012}
that can be readily utilized in quantum key distribution (QKD) \citep{Zhang2008,PhysRevA.90.062331,1367-2630-17-2-022002}.
On the other hand, their generation is challenging and full characterization,
i.e. reconstruction of the entire density matrix, requires a tremendous
effort as the number of settings of a hypothetical projective measurement
grows as $d^4$ with state dimension $d$ \citep{1751-8121-47-42-424005,Erker2017quantifyinghigh}.
This highlights the need for other characterization techniques based
on compressive sensing \citep{PhysRevLett.112.253602}, witnesses \citep{Walborn2009, Kogias2015, Deng2017}
or partial state reconstruction through semi-definite programming \citep{PhysRevLett.118.110501}. 

For bipartite HD entangled states the uncertainty relations have
proven to be a viable way of detecting entanglement \citep{Reid1989a,Walborn2009}.
The entropic steering inequalities, that stem from
the uncertainty relation of Bia{\l}ynicki-Birula and Mycielski \citep{Biaynicki-Birula1975},
are able to detect EPR-steering \textendash{} a special case of entanglement
related to the famous Einstein-Podolsky-Rosen (EPR) paradox \citep{Einstein1935a}.
In an EPR-steering experiment one party (A) is able to remotely steer the conditional state of another party (B) by selecting a measurement basis, such as position or momentum of a particle. Today, the concept of EPR-steering greatly benefits from operational definition
proposed by Wiseman \textit{et al.} \citep{Wiseman2007a} which formulates
it as a secure QKD task \citep{Branciard2012}
of one-sided device-independent QKD (1SDI-QKD). Then the maximum
achievable secure quantum key rate (per photon pair) $\Delta I$ in such a protocol can
be lower-bounded by \citep{Walborn2011}
\begin{eqnarray}
\Delta I & \geq & \mathrm{{log}(\pi e)-[\underbrace{h(x^{\mathrm{B}}\mid x^{\mathrm{A}})+h(k_{x}^{\mathrm{B}}\mid k_{x}^{\mathrm{A}})}_{\Sigma_{h}}],}\label{eq:entr-roznicz}
\end{eqnarray}
where $h(x^{\mathrm{B}}|x^{\mathrm{A}})$ and $h(k_{x}^{\mathrm{B}}|k_{x}^{\mathrm{A}})$
are conditional differential Shannon entropies for measurement in
two conjugate, e.g. position and momentum bases. An estimated positive key rate certifies EPR-steering. Henceforth we use
the binary logarithm and thus the entropy is expressed in bits. Another commonly used witness for EPR-steering, apart from the conditional entropies sum $\Sigma_{h}$ [see Eq. (\ref{eq:entr-roznicz})],
is the product of variances \citep{Reid1989a,Mancini2002a}:
\begin{equation}
\Pi_{\Delta^{2}}=\langle\Delta^{2}(x^{\mathrm{B}}-x^{\mathrm{A}})\rangle\langle\Delta^{2}(k_{x}^{\mathrm{B}}+k_{x}^{\mathrm{A}})\rangle,\label{eq:Reid-1}
\end{equation}
which certifies EPR-steering if $\Pi_{\Delta^{2}}<1/4$. This witness
is well-suited to Gaussian states as any EPR-steerable
Gaussian state is properly characterized by the variance-based witness,
i.e. it will always indicate EPR-steering when steering is present.
In consequence, it has been widely used in demonstrations of entanglement
and EPR-steering in continuous-variable systems such as quadratures
of light or collective spin \citep{Ou1992a,Takei2006a,Jensen2011}.
This translates well to QKD where it has been shown
that only Gaussian attacks need to be considered to prove security
\citep{Grosshans2004}. Although, Eq. (\ref{eq:entr-roznicz}) is more inclusive
than Eq. (\ref{eq:Reid-1}) in a sense that for a broad range of EPR-steerable
states Eq. (\ref{eq:entr-roznicz}) confirms the EPR-steering while
Eq. (\ref{eq:Reid-1}) criterion might not \citep{Walborn2009,Walborn2011,Tasca2013}.

In this Letter we demonstrate generation and characterization of an
HD entangled and EPR-steerable state of a single photon (A) and a collective
atomic excitation stored in a cold rubidium ensemble, retrievable after quantum memory storage time as another single photon (B). The HD character
relies on a wide space of angular emission modes and the conjugate
position-space of the atomic ensemble \citep{Dabrowski2017a,Chrapkiewicz2017,Parniak2017}.
The HD EPR-steering is demonstrated for the first time using a single-photon
resolving camera, which offers parallel access to all coincidences
and is thus necessary to truly utilize the high informational capacity
of the generated state \citep{Zhang2008,Leach2012d}. Furthermore,
we show that for the emerging domain of HD entanglement in position-momentum
\citep{Howell2004,Edgar2012,Moreau2014} (but also time-frequency \citep{Cho2014,Kues2017,Tiranov2017} or optical angle-orbital angular momentum (OAM) \citep{Leach2010,Zhou2015,Inoue2009,Shi2018}) 
variables of single photons the entropic witness vastly outperforms
the variance-based witness and consequently constitutes a promising way of
demonstrating HD entanglement in real systems. Finally, we use a recent result of Schneeloch and Howland \citep{Schneeloch2017}
to estimate the entanglement of formation (EOF) of the generated quantum
state. The integration
of the quantum memory presented here with a practical QKD scheme provides an
instantaneous advantage as the receiver
(Bob) could perform measurements after the sender (Alice) announces
her basis selection thus effectively doubling the key rate generation. A quantum memory also opens new avenues to fundamental studies of one-particle uncertainty principles \citep{Berta2010a}.

\section{Certifiability of EPR-steering}

The character of the HD entangled state of a photon pair differs
significantly from the one of quadrature squeezed states of light
\citep{Jensen2011}, traditionally understood as continuous-variable
states. Even if the photon pair is prepared in a Gaussian-shaped state \citep{Edgar2012}
\begin{equation}
\psi(x^{\mathrm{A}},x^{\mathrm{B}})=\frac{\kappa}{\pi\sigma}\exp\left(-\frac{(x^{\mathrm{B}}-x^{\mathrm{A}})^{2}}{4\sigma^{2}}-\frac{\kappa^{2}(x^{\mathrm{B}}+x^{\mathrm{A}})^{2}}{4}\right),\label{eq:psi}
\end{equation}
where $\sigma$ and $\kappa^{-1}$  are the s.d. of the two Gaussians, the processes of transferring the state through a quantum channel,
 detection or the multi-photon pairs will yield an accidental coincidence
background. The coincidence distribution is well-described
by 
\begin{equation}
P(x^{\mathrm{A}},x^{\mathrm{B}})=q|\psi(x^{\mathrm{A}},x^{\mathrm{B}})|^{2}+(1-q)p(x^{\mathrm{A}})p(x^{\mathrm{B}}),\label{eq:prob}
\end{equation}
 where $p(x)$ is the marginal distribution of photon counts and $0\le q\le1$ is the fraction of correlated coincidences. This
coincidence distribution is non-Gaussian unless the width $\sigma=\kappa^{-1}$ or q=1,
as the $p(x^{\mathrm{A}})p(x^{\mathrm{B}})$ term is very wide in
the $x^{\mathrm{B}}-x^{\mathrm{A}}$ variable. An analogous scenario
will be observed also for the conjugate variable \textendash{} the
wavenumber $k_{x}^{\mathrm{A}}$ and $k_{x}^{\mathrm{B}}$ in our
case. As a result, the variance-based witness immediately fails
as even for large $q$ the huge variance of wide distribution will
make the net variance large. However, the entropic witness does not
share this impairment as it goes beyond the second moments of the
measured distribution.

Figure 1 depicts these witnesses for one-dimensional situation as a function
of the coincidence fraction $q$ and the Schmidt mode number $K=(\sigma\kappa+(\sigma\kappa)^{-1})/2$
\citep{Law2004}. Note the entropic witness is easier to achieve with the
larger mode number $K$, whereas the variance-based witness exhibits
an opposite, inadequate behavior. While it has been previously demonstrated
that the entropic witness is more effective for a random class of
states \citep{Walborn2009} or Gauss-Hermite  modes \citep{Walborn2011},
here we show that for a wide variety of states present in real
experimental scenarios the entropic witness is more effective. This
leads us to a recent theoretical proposal showing that
the value $E$ of EOF, i.e. the number
of two-qubit maximally-entangled states needed to reproduce the observed
correlation \citep{Schneeloch2017,PhysRevLett.118.110501}, can be lower-bounded by the entropic witness $E\geq\mathrm{{log}(2\pi)-\Sigma_{h}}.$
The method provides an effective way of quantifying entanglement,
however here we show that the presence of accidental counts
must be taken into account.
\begin{figure}[!t]
\includegraphics[width=1\columnwidth]{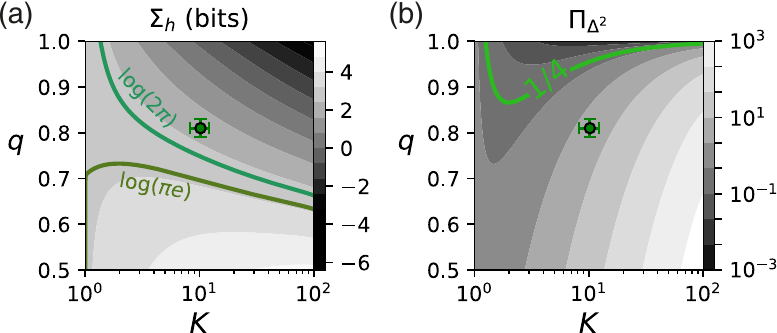}\caption{Values of the entropic (a) and variance-based (b) witnesses as a function
of Schmidt mode number $K$ and correlated coincidence fraction q. The denoted
contour bounds correspond to EPR-steering in the case of $\log(\pi e)$
for $\Sigma_{h}$ and 1/4 for $\Pi_{\Delta^{2}}$, and the demonstration
of $E>0$ in the case of the $\log(2\pi)$ bound. The dot with errorbars shows
our experimental result for the zero quantum memory storage time.
Note that since detection resolution of our experiment is 
higher than width of correlations we may project the Shannon entropy
result onto differential entropies, obtaining the same degree of violation
of respective inequalities. In both cases lighter color corresponds
to a higher value of the respective witness and consequently less
entanglement certified in the system.}
\end{figure}
\section{Experimental setup}

In order to experimentally demonstrate genuine violation, we use a single-photon
resolving I-sCMOS camera \citep{Chrapkiewicz2014,Jachura:2016aa} (see Appendix A for details) to
capture all photons emitted from a cold atomic quantum memory.
Note that previous demonstrations of HD entanglement heavily relied
on subtraction of accidental background \citep{Edgar2012,Moreau2014,Dabrowski2017a}
and in consequence only estimated the number $K$ of Schmidt modes 
potentially available while not confirming the presence of certified
entanglement or EPR-steering. Other experiments utilize various
types of spatial modes but allowed measurement of only a single mode-pair
at-a-time \citep{Howell2004,OSullivan-Hale2005a,Schneeloch2013a,Zhang:2016aa,Ding:2016aa,PhysRevLett.117.250501}.
For certification of EPR-steering we use a coarse-grained analogue
of the entropic witness \citep{Schneeloch2013a,Tasca2013}: 
\begin{eqnarray}
\Delta I & \geq & \mathrm{{log}}(\frac{\pi e}{\Delta_{x}\Delta_{k}})-[\underbrace{H(x^{\mathrm{B}}\mid x^{\mathrm{A}})+H(k_{x}^{\mathrm{B}}\mid k_{x}^{\mathrm{A}})}_{\Sigma_{H}}],\label{eq:entr-shannona}
\end{eqnarray}
where now $H(x^{\mathrm{B}}|x^{\mathrm{A}})$ and $H(k_{x}^{\mathrm{B}}|k_{x}^{\mathrm{A}})$
are conditional discrete Shannon entropies.
Note that in our case the coarse-graining resolutions $\Delta_{x}$,\ $\Delta_{k}$ are high enough
($\Delta_{x}<\sigma,\ \Delta_{k}<\kappa$) to well resolve the observed
correlations thus the differential criterion {[}Eq. (\ref{eq:entr-roznicz}){]}
is simply a shifted version of Eq. (\ref{eq:entr-shannona}) in which
the bounds remain at specified levels so any kind of offset value
such as $\Delta I$ remains unchanged.

\begin{figure}[!t]
\includegraphics[scale=0.99]{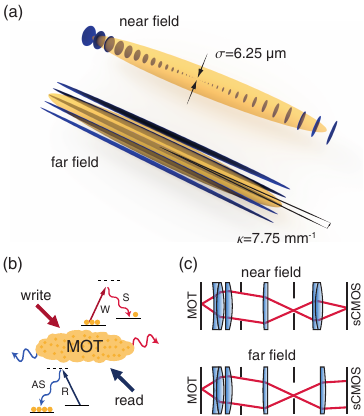}\centering\caption{(a) Representation of spin-wave modes in the atomic ensemble for which
the projective measurement is performed. While the far-field mode
is close to a transverse plane wave, the near field mode is shaped
like a tightly-focused Gaussian beam, having a wide transverse profile
at the edges of the ensemble. This highlights the importance of the
perfect overlap of detection modes determined by two separate systems in the
longitudinal direction. (b) Experimental setup: strong
write (W) laser beam produces Stokes (S) photon as a result of Raman
scattering in cold \protect\textsuperscript{87}Rb ensemble (MOT).
After adjustable storage time anti-Stokes (AS) photon is scattered
during readout (R) process. (c) Far and near field imaging setups
with iris diaphragms to select observation regions, determining the
system instrumental response and the number $K$ of observed modes. The near and far field setups
differ solely by the last lens, the aperture of which is large enough
to project all remaining modes onto the camera. \label{fig:expSetup}}
\end{figure}

The EPR-steering generation between Stokes (S) photons stored in the
quantum memory and collective atomic excitations retrieved in the form
of scattered anti-Stokes (AS) photons relies on the off-resonant Raman
scattering in a $\Lambda$-type system (see Fig. \ref{fig:expSetup}).
The experimental setup is based on the cold rubidium-87 ensemble inside a magneto-optical
trap (MOT) (see Appendix B for details). The scattered
photons are imaged on the sCMOS camera equipped with an image intensifier \citep{Chrapkiewicz2014}.
Both near (position basis) and far field (momentum basis) imaging
setups consist of four lenses: near field with magnification $M=8$ and far field with effective focal length $f_{\mathrm{eff}}=50\:\mathrm{{mm}}$.
High resolution of the near field imaging with two separate systems
for S and AS photons is achieved by overlapping the focal points of
the two systems in the longitudinal direction within the Rayleigh
range of 38 $\mu\mathrm{{m}}$. Write pulse contains $10^{6}$ photons, resulting in registering on
average $\bar{n}=0.1$ Stokes photon per camera frame. We estimate memory
readout efficiency of 35\%, taken as the ratio of coincidence rate to S photons rate and accounted for losses \citep{Parniak2017}.

In our experimental setup the coarse-grained elementary cells \citep{Walborn2009,Tasca2013}
have the size of $\Delta_{x}={\delta_{b}}/M=3.88\:\mu\mathrm{m}$
($\Delta_{k}={2\pi\delta_{b}}/(\lambda f_{\mathrm{eff}})=4.86\:\mathrm{mm^{-1}}$)
for near (far) field, where $\delta_{b}=31\ \mu\mathrm{m}$ is the physical bin size on the detector plane and $\lambda$
is scattered photon wavelength (equal either 780 nm or 795 nm). The experimental uncertainties of
the magnification $M$ and focal length $f_{\mathrm{eff}}$ 
determine the uncertainty of the right-hand-side of Eq. (\ref{eq:entr-shannona}),
increasing our errorbar on the attainable key rate $\Delta I$ or
entanglement of formation $E$. On the camera we observe S and AS photons in separate
circular regions of $100\delta_b$ diameter corresponding to an image
of the circular region of $0.4\:\mathrm{mm}$ diameter for the near
field ($62\ \mathrm{mrad}$ scattering angle for the far field). These
regions are selected using iris diaphragms situated in intermediate
image and Fourier planes in the part of the imaging setup common
to both systems {(}see Fig. \ref{fig:expSetup}(c){)}.
\section{Variance-based witness}

\begin{figure}[!t]
\includegraphics[scale=0.52]{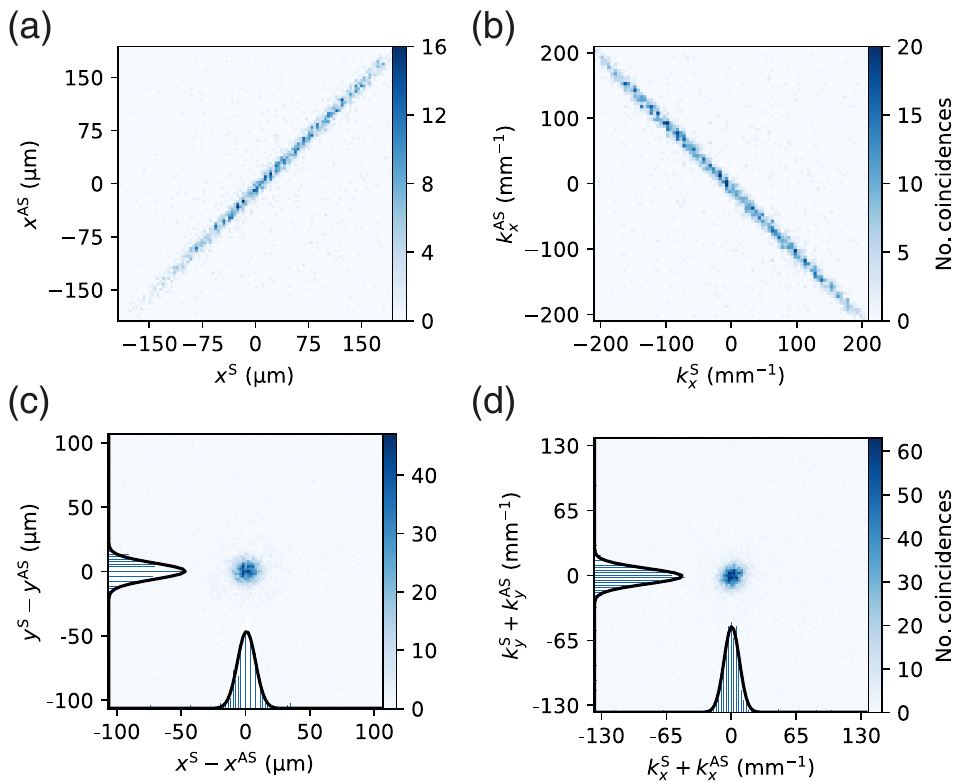}\centering\caption{Joint coincidence distributions for zero memory storage time for position
(a) and momentum (b) basis in the $x$-dimension and the difference of position (c) and sum of momenta (d) for the entire
two-dimensional distribution. Gaussian functions are
fitted to the cross-sections through the distributions maxima to obtain
modes size $\sigma$ ($\kappa$) in the near (far)
field. Each measurement comprises $10^7$ camera frames containing approx. 3000 coincidences.\label{fig:corrMaps}}
\end{figure}

In Fig. \ref{fig:corrMaps}(a) and  \ref{fig:corrMaps}(b) we present joint coincidence
distributions of scattered Stokes\textendash anti-Stokes pairs for
positions $(x^{\mathrm{S}},x^{\mathrm{AS}})$ and momenta $(k_{x}^{\mathrm{S}},k_{x}^{\mathrm{AS}})$
respectively, registered for zero memory storage time. In Fig. \ref{fig:corrMaps}(a) we post-select from full four-dimensional distribution $n(x^\mathrm{A},y^\mathrm{A},x^\mathrm{B},y^\mathrm{B} )$ the detection events with $y^\mathrm{AS}=y^\mathrm{S}\pm2\sigma$ (and $k_y^\mathrm{AS}=-k_y^\mathrm{S}\pm2\kappa$ from ($k_x^\mathrm{A},k_y^\mathrm{A},k_x^\mathrm{B},k_y^\mathrm{B}$ ) for Fig. \ref{fig:corrMaps}(b) where $\sigma$($\kappa$) is the size of the spatial mode in the near (far) field (inferred from Fig. \ref{fig:corrMaps}(c) and  \ref{fig:corrMaps}(d), thus the $y$-dimension
can be treated as a herald of coincidence event in $x$-direction.
Assuming both $n(x^\mathrm{A},x^\mathrm{B})$ and $n(k_x^\mathrm{A},k_x^\mathrm{B})$ are described in terms of Gaussian function (see Eq. (\ref{eq:psi}) and its Fourier transform), we estimate the number of post-selected independent spatial modes for one-dimensional situation equal $K=10.2$ (ref. \citep{Law2004}) which corresponds to $K_{2\mathrm{D}}\simeq K^2\approx104$
modes for two dimensions. Note that considering only the near or far field (comparing
long and short axes of the observed Gaussian-shaped distribution \citep{Parniak2017} -- see Eq. (\ref{eq:psi})) we obtain slightly
higher estimate of the number of modes $K\approx14$ that can be used e.g., for multiplexing purpose. The slightly smaller number of modes estimated with the first method than the number of modes inferred from a single coincidence distribution comes from residual misalignment of near and far fields in the experimental setup. The two bases (near and far field) are not perfectly conjugate and the size of the spatial mode is larger in one field compared to the situation with perfect alignment. The observed state thus slightly deviates from a pure state given by Eq. \ref{eq:psi} and is better described by a density matrix, yet the difference is small and we choose to use the more concise pure-state description.

Figures \ref{fig:corrMaps}(c)
and  \ref{fig:corrMaps}(d) are inferred from the same data but for different choices
of basis vectors, namely $(x^{\mathrm{S}}-x^{\mathrm{AS}},\:y^{\mathrm{S}}-y^{\mathrm{AS}})$
and $(k_{x}^{\mathrm{S}}+k_{x}^{\mathrm{AS}},\:k_{y}^{\mathrm{S}}+k_{y}^{\mathrm{AS}})$.
The cross-sections through the peak maximum are used to infer the
value of variance-based witness $\Pi_{\Delta^{2}}$ {[}Eq. (\ref{eq:Reid-1}){]}.
By fitting a Gaussian function to only the central peak we recover
the mode widths $\sigma=6.25\ \mu\mathrm{{m}}$ and $\kappa=7.75\ \mathrm{{mm^{-1}}}$ {(}see Fig. \ref{fig:expSetup}(a){)}.
We immediately find that while the expected value for the noiseless
scenario of the variance-based witness is $\Pi^{\mathrm{(expected)}}_{\Delta^{2}}=\sigma^{2}\kappa^{2}=2.4(2)\times10^{-3}$,
the actually observed value is $\Pi_{\Delta^{2}}=7(1)$ \textendash{}
far from certifying EPR-steering. The uncertainties of inferred variances
are combinations of standard deviations calculated using Monte Carlo
method and experimental uncertainties. The above result shows that EPR-steering
criterion based on variances is ill-suited for the state we generate
in the experiment thus positive secure key rate could not be established
just by monitoring the variances \citep{Grosshans2004}.

\section{Shannon entropy-based witness}
\begin{figure}[!t]
\includegraphics[width=0.85\columnwidth]{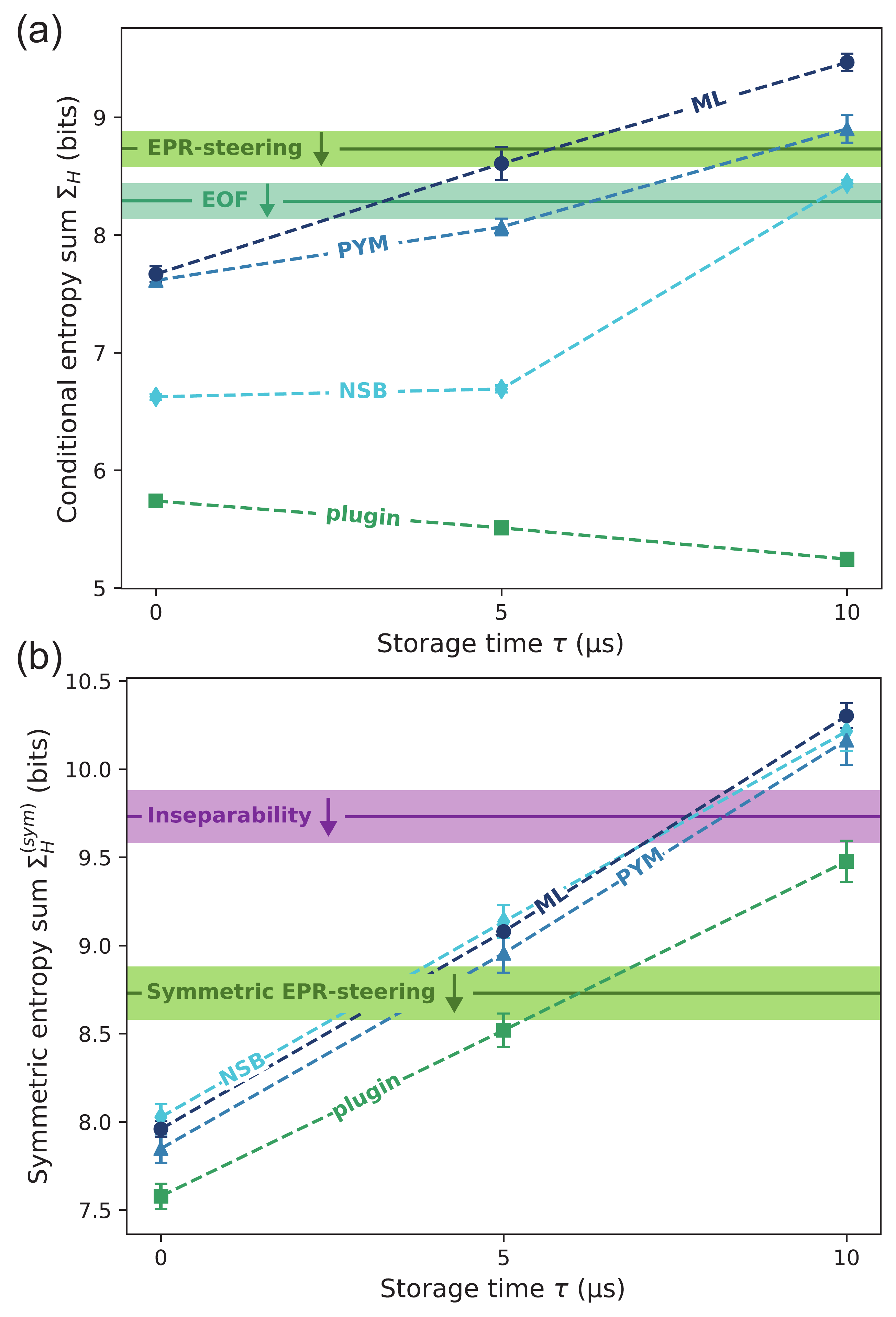}\centering\caption{Panel (a). Conditional entropy sum $\Sigma_{H}$ inferred from experimental data for different memory storage
times. Subsequent series of data (from bottom) correspond to: na\"ive (``plugin''),
Nemenman-Shafee-Bialek (NSB) \citep{Nemenman2004}, Pitman-Yor mixture
(PYM) \citep{Archer2014} and parametric maximum likelihood
(ML) estimators of Shannon's entropy, respectively. Panel (b). Symmetric entropy sum $\Sigma_H^{\mathrm{(sym)}}$ for certification of symmetric EPR-steering and inseparability. For the Bayesian estimators the uncertainties are inferred from variances of posterior distributions, for the parametric ML estimator from a Fisher information matrix, while for ``plugin'' are calculated using Monte Carlo approach. Limits for EPR-steering
$\Delta I\geq0$ and ceritifed entanglement of formation (EOF) $E\geq0$
are marked. Shaded regions around the respective limits stem from
uncertainties of imaging system calibration. Values below the respective
limits indicate that the generated state resides deeper in the quantum
regime. \label{fig:entropyEst}}
\end{figure}

Estimating entropy of sampled probability distribution from experimental
data has been shown to be a hard task, as no unbiased non-parametric
estimator exists \citep{Paninski2003}. Here we infer the conditional entropy values from the experimental data (Fig. \ref{fig:corrMaps}) as a difference of joint and
marginal entropy distributions $H(x^{\mathrm{B}}|x^{\mathrm{{A}}})=H(x^{\mathrm{B}},x^{\mathrm{{A}}})-H(x^{\mathrm{{A}}})$.
Figure \ref{fig:entropyEst}(a) presents the values of the entropic witness
$\Sigma_{H}$ [Eq. (\ref{eq:entr-shannona})], inferred using four different Shannon entropy estimators
\citep{Archer2014} (see Appendix C for details). A na\"ive ``plugin'' estimator,
based on relative frequencies directly plugged into the formula
for Shannon entropy as probabilities, is strongly biased towards lower
entropy values, unless the total number of counts exceeds the total
number of available bins by orders of magnitude. In the case of EPR-steering verification this leads to the
Eq. (\ref{eq:entr-shannona}) being falsely satisfied. In particular,
$\Sigma_{H}$ estimated using the na\"ive method drops with the storage
time $\tau$ as an effect of falling number of counts due to the quantum memory
retrieval efficiency decay, which in turn results in a higher negative
bias.
This crucial problem has been addressed by Bayesian estimation
through applying prior mixtures highly non-informative about the entropy: a Dirichlet
mixture (NSB estimator) \citep{Nemenman2004} and the Pitman-Yor mixture
(PYM estimator) \citep{Archer2014}. These estimators are specifically
designed to deal with undersampled number of counts. Additionally,
we construct a parametric estimator that assumes a Gaussian shape
of the distribution on top of a wide accidental background. The coincident counts
probability distribution is fitted to experimental data by maximizing
the likelihood function and the entropy is calculated directly
from the fitted distribution (ML estimator).

The consistency of ML and PYM estimators is a strong indication
that the more general PYM estimator can be considered trustworthy
in our particular scenario. On the other hand, NSB estimator shows
a constant bias for all measurements while still performing better
than the ``plugin'' estimator. The parametric ML and PYM estimators
agreeably certify EPR-steering for $\tau=0$ as well as
for the $\tau=5\ \mu\mathrm{{s}}$, but fail for the case of $\tau=10\ \mu\mathrm{{s}}$ in which the secure information capacity $\Delta I$
drops to zero as accidental coincidences dominate over the Stokes\textendash anti-Stokes
coincidences. The two main limiting factors are the decay of memory
retrieval efficiency due to motional dephasing and the increasing
noise due to influx of untrapped atoms from the residual gas into
the interaction region within the readout beam. Nevertheless, for $\tau=0$ we obtain $\Delta I=1.06\pm0.15$ bits per photon.
We also certify entanglement of formation $E\geq0.70\pm0.15$ ebits. We stress
that even for Gaussian states not all EOF is
certified with this method \citep{Schneeloch2017}, and for the state
generated here more ebits are likely needed to describe correlations. The
number of modes $K$ places an upper limit on
EOF equal $E\leq\log(K)\approx3.3$ ebits, however currently there is no method to better estimate the EOF. 

Nevertheless, we may still certify the inseparability of the bipartite system without referring to the amount of entanglement in terms of EOF. As the class of entangled states contains all EPR-steerable states \citep{Wiseman2007a}, the witness based on conditional entropies certifies entanglement if $\Delta I\geq 0$, similarly as for EPR-steering. The conditional entropies cannot form a more inclusive witness; however, we may construct a slightly modified witness based on entropies of probability distributions for $(x^\mathrm{B}-x^\mathrm{A})$ and  $(k_x^\mathrm{B}+k_x^\mathrm{A})$  variables \citep{Walborn2009}:
\begin{equation}
\Sigma_H^{\mathrm{(sym)}}=H(x^\mathrm{B}-x^\mathrm{A} )+H(k_x^\mathrm{B}+k_x^\mathrm{A} )
\end{equation}
Such witness offers two distinct features: if $\Sigma_H^{\mathrm{(sym)}}\leq \log(2\pi e/\Delta_x \Delta_k)$ the state is inseparable, and if $\Sigma_H^{(sym)}\leq \log(\pi e/\Delta_x \Delta_k)$, the state is EPR-steerable in both directions. The latter condition is slightly more exclusive than $\Delta I \geq 0$ [Eq. (5)], but the inseparability can be certified more easily than in the asymmetric (conditional) case.  In Fig. \ref{fig:entropyEst}(b) we plot the experimentally inferred values of the symmetric witness along with the two bounds. We clearly observe that inseparability is unambiguously certified for storage times of up to 5 $\mu$s, with $1.77\pm0.15$ bits of difference between $\Sigma_H^{(sym)}$ and $\log(2\pi e/\Delta_x \Delta_k)$ for parametric ML estimator (for $\tau=0$). Note that in the asymmetric case presented in Fig. \ref{fig:entropyEst}(a) the condition for inseparability overlaps with the condition for EPR-steering.

\section{Discussion}
In conclusion, we presented the generation and storage of an atom-photon
HD EPR-steerable state without noise subtraction. We showed,
using entropic steering inequality, that the atomic part of the generated
state can be stored up to 5 $\mu\mathrm{{s}}$, long enough to perform
electronic feedback-loop based on the results of Stokes photon measurements
\citep{Parniak2017}. With such system an active multiplexing technique
could be employed \citep{Mazelanik2016} to obtain a one-dimensional stripe of entangled modes
with a high correlated coincidence fraction $q$. We also demonstrate
the superiority of the entropic entanglement witness \textendash{}
even if the variance EPR-steering inequality (Eq.~(\ref{eq:Reid-1}), \citep{Reid1989a}) is not
satisfied, the corresponding entropic version (Eq.~(\ref{eq:entr-roznicz}), \citep{Tasca2013,Schneeloch2013a}) confirms EPR-steering. We highlighted and solved the difficulties in entropy estimation, achieving characterization of the HD entangled state with highly undersampled data. 

We envisage that with higher numbers of modes $K$ the entropic witness [Eq. (\ref{eq:entr-roznicz})]
may detect EPR-steering even without the detection efficiency loophole.
As the value of $q$ required to witness EPR-steering drops with
$K$, loophole-free demonstration could be achieved for large $K$
even with low quantum efficiency of the detector \citep{Vallone2013}. Moreover, in a practical
scenario a secure transmission could be established over a very lossy
channel. This would be an essential step towards HD free-space QKD, particularly in space applications \citep{PhysRevLett.113.060503,Yin2017}.

\begin{acknowledgments}
We thank I. Bia\l ynicki-Birula and R.~Demkowicz-Dobrza\'{n}ski
for discussions as well as K. Banaszek for generous support. This project has been financed by the National Science Center (Poland) (Grants No. 2015/19/N/ST2/01671, 2016/21/B/ST2/02559, 2017/25/N/ST2/01163 and 2017/25/N/ST2/00713) and Polish MNiSW Diamentowy Grant (Projects No. DI2013 011943 and DI2016 014846). M. D. is supported by the Foundation for Polish Science (FNP).
\end{acknowledgments}

\appendix

\section{Detection via I-sCMOS camera}
\begin{figure}[b]
\includegraphics[scale=1.7]{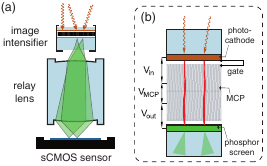}\centering \caption{(a) Scheme of the I-sCMOS camera used to collect single photons. (b) Image intensifier module. Photons impinging on the photocathode produce avalanches of electrons inside micro-channel plate (MCP) which results in phosphor flashes on the screen. See text for details. \label{fig:I-sCMOS_scheme}}
\end{figure}
To localize photons with high spatial resolution we use an I-sCMOS camera system consisting of scientific complementary metal-oxide semiconductor (sCMOS) sensor endowed with image intensifier, assembled in our group from commercially available components. The sketch of the I-sCMOS device is depicted in Fig. \ref{fig:I-sCMOS_scheme}(a). 

A third generation image intensifier (Hamamatsu V7090D), powered by an external gating module consists of a photocathode, a micro-channel plate (MCP) and a phosphor screen. The image intensifier power module (Photek, GM300-3), enabling gating down to 5 ns, is gated synchronously with drive laser pulses using the FPGA. High voltage is delivered by a separate module (Photek, FP830). Photons impinging through the input window on the 18 mm diameter gallium arsenide (GaAs) photocathode are converted to photoelectrons with 20\% quantum efficiency. The details of image intensifier module are presented in Fig \ref{fig:I-sCMOS_scheme}(b).

Photoelectrons are attracted from photocathode to MCP by short pulses of gate voltage $V_\mathrm{in}=-200$ V. Normally the voltage $V_\mathrm{in}=+50$ V and turns off the image intensifier. Electrons reach the adjacent MCP under high voltage $0.25 \le V_{\mathrm{MCP}}\le 1.75$ kV that accelerate electrons in $\sim9\times10^6$ small, separated channels of ca. $6\ \mu$m diameter each. The channels are tilted by several degrees from the axis (and twisted in the middle to prevent direct electrons  passing) thus the electrons hit the channel walls and produce a stochastically growing charge avalanche. The typical gain, i.e. the mean multiplication factor (tuned by changing the MCP voltage $V_{\mathrm{MCP}}$) is about $10^{5}$. Electrons leaving MCP are accelerated by a high voltage $V_{\mathrm{out}}=6$ kV and hit the phosphor screen producing bright flashes of green light ($66$ $\mu$m diameter, decay time below 200 ns).

The phosphor flashes emitting light in the wide angle are subsequently imaged onto the fast, low-noise $6.5 \times 6.5\:\mathrm{mm}$ pixel size sCMOS sensor (Andor Zyla 5.5 Megapixel) by a high numerical aperture relay lens (Stanford Computer Optics, $f/\#=1.1$). The focal length $f=105$ mm of the relay lens and the magnification $M=-0.44$ result in an image size of the phosphor screen corresponding to a circle of 1230 pixels (8mm) in diameter, hence we effectively use about $1.2\times10^{6}$ camera pixels altogether. We operate the camera in the fastest shutter mode enabling the readout of the full frame with 100 fps frame rate which can be significantly increased by reducing the number of readout camera rows. This is essential for collecting large photon statistics in quantum optics experiments where one can operate with low photon numbers per frame.

The individual single photon flashes are registered as 5-pixel FWHM Gaussian spots on the sCMOS sensor. A single pixel signal is digitized to 2048 analog-to-digital units (ADU). The average pixel offset is at the level of 100 ADU and  varies less than 5 ADU across the sensor. Moreover, the s.d. of the signal when measuring dark counts is at the level of 2.5 ADU (1.5 electrons). The average intensity of the brightest pixel (highly random brightness is determined by the stochastic avalanche process) is approximately 500 times higher than the camera noise level thus the spot can be easily discriminated from the low-noise background.

A real-time software algorithm (C compiled DLL library) retrieves central positions of the single flashes from a raw image with 1.5 $\mu$m subpixel resolution and bins them into camera pixels. The central position of the spot is found by fitting a two-dimensional paraboloid to the logarithm of the pixel intensity, which must exceed the present treshold at least five times, within the three pixels radius. Fast and efficient fitting is performed by evaluating a numerical integral on the neighboring pixel intensities. The real-time algorithm reduces the amount of stored data enormously -- instead of large raster camera frames we store only 16-bit unsigned integer numbers describing the transverse positions of the photons. 

\section{Cold atoms quantum memory}
In the experimental sequence we first trap the atoms using the MOT for 1.4 ms. In the steady state we trap approximately $N=10^8$ atoms at 100 $\mu$K. Next, we turn off the trapping magnetic field and detune the cooling laser 35 MHz to the red from the $5^2 S_{1/2},F=2\to5^2 P_{3/2},F'=3$ transition. The optical molasses is maintained for 300 $\mu$s and we reach the final temperature of $22\pm2\ \mu$K and optical depth $\mathrm{OD}=40$. The final atomic ensemble is 6-mm-long and has a transverse width of $w=0.6\pm0.1$ mm (which we calculate as the $1/e^2$ diameter of the atomic column density). After the cooling and trapping stage we optically pump the atoms into a single Zeeman $F=1,m_F=1$ state \citep{Parniak2017}, which takes 40 $\mu$s. The quantization axis is set by a weak bias magnetic field of approx. 100 mG. 

The quantum memory is used once per atomic ensemble prepared with this sequence. After a short dark period of 1 $\mu$s that allows all residual excitations due to pump fields to decay we generate the bipartite entangled state by illuminating the ensemble with a 100 ns, off-resonant write pulse with wavevector $\mathbf{k}_\mathrm{w}$ (left-circular polarization, red-detuned by 20 MHz from the $5^2 S_{1/2},F=1\to5^2 P_{3/2},F'=2$ transition). As a result, a Raman-scattered Stokes (S) photon with a wavevector $\mathbf{k}_\mathrm{S}$ is generated together with a collective atomic excitation (spin wave) with a wavevector $\mathbf{K}=\mathbf{k}_\mathrm{w}-\mathbf{k}_\mathrm{S}$. After quantum memory storage time, the spin wave is converted on-demand to an anti-Stokes (AS) photon through resonant Raman scattering with a 200 ns read pulse using with a wavevector $\mathbf{k}_\mathrm{r}$ (readout beam counter-propagating with respect to the write-in, right-circular polarization, resonant with the $5^2 S_{1/2},F=2\to5^2 P_{1/2},F'=2$ transition). Wavevector $\mathbf{k}_{\mathrm{AS}}$ of the AS photon is determined by the stored atomic excitation: $\mathbf{k}_{\mathrm{AS}}=\mathbf{K}+\mathbf{k}_\mathrm{r}$.

We additionally use two separate (one for write-in and the other for readout) optically-pumped hot rubidium vapour cells with buffer gas and paraffin coating \cite{Parniak2017} to obtain spatially-insensitive filtering, essential to observe spatial entanglement. The cells are pumped by strong lasers during the cooling and trapping period of the MOT. Combined average transmission of the imaging and filtering system is 40$\%$. Finally, photons originating from the atomic quantum memory are imaged onto separate regions of the same the I-sCMOS sensor (located in the image or Fourier plane) through a diffraction-limited imaging setup, as depicted in Fig. 2c.

\section{Shannon entropy estimators}
Here, we describe the method of inferring the value of Shannon entropy estimators presented in Fig. 4 from the experimental data, measured using I-sCMOS camera. In the experiment we register the positions $(x^\mathrm{A},y^\mathrm{A})$, $(x^\mathrm{B},y^\mathrm{B})$ (momenta $(k_x^\mathrm{A},k_y^\mathrm{A})$, $(k_x^\mathrm{B},k_y^\mathrm{B})$) of photons in write-in (A) and readout (B) regions of I-sCMOS camera, depending on imaging system (near or far field).

We are working in the low photon-number regime where rarely more than single count occurs either in A and B region and post-select frames where exactly one photon is register in A and one photon is registered in B. With many frames we arrive at coincidence distribution $n(x^\mathrm{A},y^\mathrm{A},x^\mathrm{B},y^\mathrm{B})$ in the near ($n(k_x^\mathrm{A},k_y^\mathrm{A},k_x^\mathrm{B},k_y^\mathrm{B})$ in the far) field. Assuming the registered coincidences are position--correlated (momenta--anti-correlated), the $y$-dimension can be treated as a herald of coincidence event in $x$-direction. Thus, we post-select the detection events with $y^{\mathrm{B}}=y^{\mathrm{A}}\pm2\sigma$
($k_{y}^{\mathrm{B}}=-k_{y}^{\mathrm{A}}\pm2\kappa$), where $\sigma$ ($\kappa$) is the size of
the spatial mode in the near (far) field, and finally obtain joint, two-dimensional distribution $n(x^\mathrm{A},x^\mathrm{B})$ ($n(k_x^\mathrm{A},k_x^\mathrm{B})$).
\begin{figure}[b!]
\includegraphics[scale=0.60]{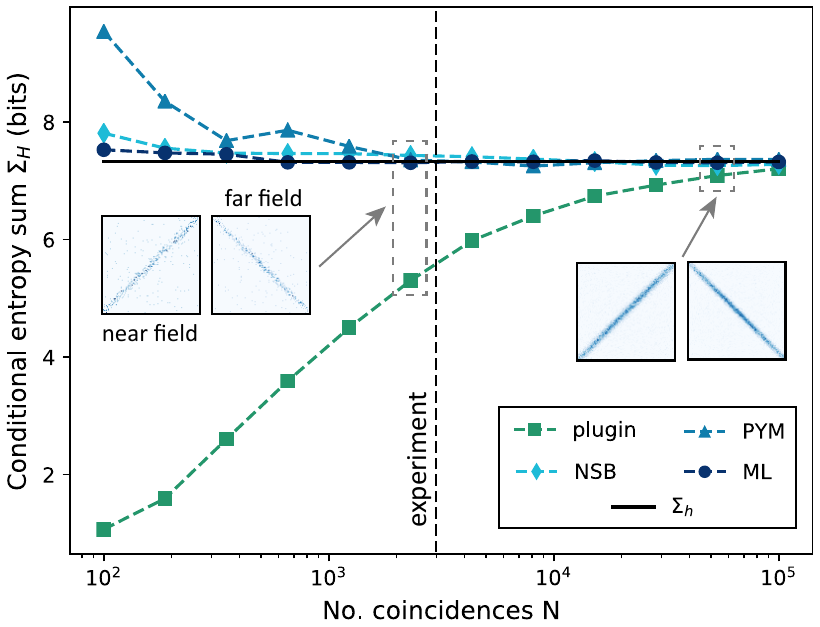}\centering \caption{Simulation of conditional entropy sum as a function of total number of coincidences. Different curves corespond to the estimators presented in Fig. 4. Dashed vertical line shows the number of coincidences registered in the experiment (approx. 3000 coincidences), while the solid horizontal line depicts the $\Sigma_{\mathrm{h}}$, calculated from probability density distribution defined in Eq. (4). Coincidence maps present the results of near and far field distributions calculated for two particular number of coincidences, from which the entropy estimators were calculated. See text for details. \label{fig:entropy_sim}}
\end{figure}
Based on joint $n(x^\mathrm{A},x^\mathrm{B})$ and marginal $n(x^\mathrm{A})$ coincidence distributions (integrated over all possible $x^\mathrm{B}$ coordinate of a second photon), we infer the conditional entropy values as a difference of joint and marginal entropy distributions $H(x^{\mathrm{B}}|x^{\mathrm{{A}}})=H(x^{\mathrm{B}},x^{\mathrm{{A}}})-H(x^{\mathrm{{A}}})$. Then we make the analogous procedure for the far field. Finally, we calculate the EPR entropy witness [Eq. (5)]: $\Sigma_H=H(x^{\mathrm{B}}|x^{\mathrm{{A}}})+H(k_x^{\mathrm{B}}|k_x^{\mathrm{{A}}})$. The details of calculation depends on specific type of estimator, as described below. For the symmetric steering witness $\Sigma_H^{\mathrm{(sym)}}=H(x^\mathrm{B}-x^\mathrm{A} )+H(k_x^\mathrm{B}+k_x^\mathrm{A} )$  we use a similar procedure by first binning the coincidences in terms of the $(x^\mathrm{B}-x^\mathrm{A})$, $(k_x^\mathrm{B}+k_x^\mathrm{A})$ variables for the near and the far field, respectively. The final distributions from which the entropies are inferred are one-dimensional, and thus the issue of undersampling is less severe for this case. In subsequent considerations we keep the two-dimensional notation as a more general case.

A na\"ive ``plugin'' estimator is
based on relative frequencies directly plugged as probabilities into the formula
for Shannon entropy:
\begin{equation}
H_\mathrm{plug}(x^{\mathrm{{A}}},x^{\mathrm{{B}}})=-\sum_{x^\mathrm{A},x^{\mathrm{{B}}}}{p(x^\mathrm{A},x^{\mathrm{{B}}})\log p(x^\mathrm{A},x^{\mathrm{{B}}})},
\label{eq_ent}\end{equation}
summed over $x^\mathrm{A}$, $x^\mathrm{B}$ (treated as discrete pixel coordinates), with probability $p(x^\mathrm{A},x^{\mathrm{{B}}})=n(x^\mathrm{A},x^{\mathrm{{B}}})/\sum{n(x^\mathrm{A},x^{\mathrm{{B}}})}$. Figure \ref{fig:entropy_sim} presents the results of entropy sum $\Sigma_H$ simulation for different number of coincidences $N$. We randomly choose $N$ events from the probability distributions defined by Eq. (4) for the Gaussian-shaped state [Eq. (3)] and its Fourier transform. We assume the same $\sigma,\:\kappa$ and $q$ as in the real experiment and calculate the probability distribution [Eq. (4)] using $100\times100$ square grid in both near and far field. Figure \ref{fig:entropy_sim} shows obtained distributions for two particular number of coincidences as well as mean results for the conditional entropy sum $\Sigma_H$ obtained using different entropy estimators, averaged over 10 different realizations. Additionally, we plot the conditional differential entropy sum $\Sigma_h$ from Eq. (1), corresponding to the asymptotic value when $N\rightarrow\infty$.

The ``plugin'' estimator is strongly biased towards lower
entropy values, unless the total number of coincidences exceeds the total
number of available bins by orders of magnitude. This crucial problem has been addressed by calculating a Dirichlet prior
mixture (NSB estimator) \citep{Nemenman2004} and the Pitman-Yor prior mixture
(PYM estimator) \citep{Archer2014}, following the formulas in the respective articles. These estimators are specifically designed to deal with undersampled number of coincidences and we can see in Fig. \ref{fig:entropy_sim} they converge to $\Sigma_h$ even for relatively small number of coincidences.

Finally,
we construct a parametric ML estimator assuming a Gaussian-shaped
distribution on top of a wide, flat accidental background: $\lambda(x^{\mathrm{A}},x^{\mathrm{B}})=\alpha|\psi(x^{\mathrm{A}},x^{\mathrm{B}})|^{2}+\beta$, where $\psi(x^{\mathrm{A}},x^{\mathrm{B}})$ is defined in Eq. (3). The analogous formula is obtained for the far field using Fourier transform of $\psi(x^{\mathrm{A}},x^{\mathrm{B}})$. The coincidence
probability distribution has been fitted to experimental or simulated data by maximizing
the likelihood function $\langle\log f(n,\lambda)\rangle$ over $\alpha,\:\beta$, where the avereging is carried over all values of $x$ or $k$, with $f(n,\lambda)=\lambda^ne^{-\lambda}/n!$ being Poissonian distribution. The entropy is calculated directly
from the fitted distribution $\lambda$ using Eq. (\ref{eq_ent}).

\bibliographystyle{apsrev4-1}

\end{document}